\documentstyle[12pt,amsmath,cite]{article}

\evensidemargin=\oddsidemargin
\newcommand{\be}{\begin{equation}}
\newcommand{\ee}{\end{equation}}
\newcommand{\bea}{\begin{eqnarray}}
\newcommand{\eea}{\end{eqnarray}}
\newcommand{\bs}{\boldsymbol}

\newcommand{\p}{\partial}

\def\half{\frac{1}{2}}

\def\IB{\relax\hbox{$\inbar\kern-.3em{\rm B}$}}
\def\IC{\relax\hbox{$\inbar\kern-.3em{\rm C}$}}
\def\ID{\relax\hbox{$\inbar\kern-.3em{\rm D}$}}
\def\IE{\relax\hbox{$\inbar\kern-.3em{\rm E}$}}
\def\IF{\relax\hbox{$\inbar\kern-.3em{\rm F}$}}
\def\IG{\relax\hbox{$\inbar\kern-.3em{\rm G}$}}
\def\IGa{\relax\hbox{${\rm I}\kern-.18em\Gamma$}}
\def\IH{\relax{\rm I\kern-.18em H}}
\def\IK{\relax{\rm I\kern-.18em K}}
\def\IL{\relax{\rm I\kern-.18em L}}
\def\IP{\relax{\rm I\kern-.18em P}}
\def\IR{\relax{\rm I\kern-.18em R}}
\def\IZ{\relax{\rm Z\kern-.5em Z}}

\begin{document}

\begin{titlepage}

\begin{flushright}
SNUTP/05-017\\
hep-th/0511205\\
\end{flushright}

\vskip .2in

\begin{center}

{\large\bf Deformed conformal and super-Poincar\'e symmetries in the non-(anti)commutative spaces}

\vskip .3in

{\bf  Rabin Banerjee}\footnote{e-mail: {rabin@bose.res.in}},

\vskip .2in

{\it S N Bose National Centre for Basic Sciences,\\
J D Block, Sector III, Salt Lake, Kolkata 700\,098, India\\
\vskip .2in
{\bf  Choonkyu Lee}\footnote{e-mail: {cklee@phya.snu.ac.kr}} 
and 
{\bf Sanjay Siwach}\footnote{e-mail: {sksiwach@phya.snu.ac.kr}}

\vskip .2in

{\it School of Physics and Centre for Theoretical Physics,\\
 Seoul National University,}
 Seoul,  151-747, Korea }

\vskip .5in

\begin{abstract}

\vskip .1in

Generators of the super-Poincar\'e algebra in the non-(anti)commutative superspace are represented using appropriate higher-derivative operators defined in this quantum superspace. Also discussed are the analogous representations of the conformal and superconformal symmetry generators in the deformed spaces. This construction is obtained by generalizing the recent work of Wess et al on the Poincar\'e generators in the $\theta$-deformed Minkowski space, or by using the substitution rules we derived on the basis of the phase-space structures of non-(anti)commutative-space variables. Even with the nonzero deformation parameters the algebras remain unchanged although the comultiplication rules are deformed. The transformation of the fields under deformed symmetry is also discussed. Our construction can be used for systematic developments of field theories in the deformed spaces.

\end{abstract}

\end{center}

\vfill

\end{titlepage}

\section{Introduction}

In present-day physics, noncommutativity of coordinates \cite{Snyder} is no longer taken as something unusual. It occurs for instance when one studies the charged particle dynamics in a strong magnetic field with lowest Landau-level projection \cite{Dunne:1989hv}. It also arises in string theory with a non zero B-field and in certain limit the theory reduces to a gauge theory on a noncommutative space \cite{Seiberg:1999vs}. Many interesting features of non-commutative field theories, e.g., UV/IR mixing, the soliton solutions, etc. have been discussed extensively in the recent literature (for a review and the list of the references, see \cite{Douglas:2001ba}). In these developments it has been typical to promote  the ordinary field-theory Lagrangian to the Lagrangian in the non-commutative spacetime by using the Moyal-Weyl $\star$ product\cite{Moyal}. 

The non-commutativity of the spacetime coordinates is often taken as the signature for violation of Lorentz invariance. But, recently, it has been shown by using the (twisted) Hopf algebra \cite{Chaichian:2004za} that corresponding field theories possess {\it {deformed}} Lorentz invariance. This suggests above all to use the representation theory of the deformed Poincar\'e algebra as a basis for systematic field theoretic discussions of these theories. In the related developments, Wess and his collaborators \cite{Wess:2003da,Dimitrijevic:2003wv} have developed the differential calculus which is more suitable for the field theory applications. Based on the latter approach, an explicit differential realization of the deformed space-time generators can be found, a novelty here being the appearance of the higher derivative terms. An application of these ideas for investigating deformed Schr\"odinger symmetry has also been made in \cite{Rabin}. In this work we show that the approach of Wess et al can be generalized to the case of the non-(anti)commutative superspace, and find explicit derivative realizations for the generators of the deformed super-Poincar\'e symmetries. Conformal and superconformal symmetries are also discussed along the same lines. A simple change-of-variable-type formula which relates these spacetime generators of the deformed space to the corresponding (more familiar) expressions of the undeformed space, is given. These results are then used to derive the appropriate field transformation laws in the deformed space and to discuss corresponding symmetries.

The plan of this paper is as follows. In the next section we will first review the formalism of \cite{Wess:2003da,Dimitrijevic:2003wv} and apply it to the conformal generators in the $\Theta$-deformed Minkowski space. As a result of deformation, these generators contain higher derivative operators, and we provide a simple understanding on these by considering the phase-space roles of noncommutative-space position and momentum variables. The comultiplication rules are also worked out using these higher-derivative realizations, and the results are consistent with those obtained through Hopf-algebraic considerations. Then, in section III, we consider  the superspace with non-(anti)commutative Grassmannian coordinates \cite{Ferrara:2000mm,Seiberg:2003yz} and develop the necessary differential calculus in this deformed superspace. With this machinery we explicitly construct the deformed generators of super-Poincar\'e group and also the superconformal generators in the form of higher-derivative operators in the deformed superspace. The very expressions also follow as we consider the appropriate change of variables with the familiar linear-differential-operator forms relevant to the undeformed superspace case. The thus constructed generators satisfy the same algebras as in the undeformed case, but give rise to certain modifications in the comultiplication rules. In section IV we discuss the transformation properties of the fields under conformal and super-Poincar\'e group. The field transformation laws, obtained using appropriate higher-derivative representations of the previous sections, give a manifestly invariant meaning to the equations of motion in the deformed space; for this, however, the transformation laws for the (local) composite fields must be introduced in accordance with the comultiplication rules. The section V is the concluding remark.

\section{Deformed Minkowski-space calculus and conformal algebra}

Consider the $\Theta$-deformed Minkowski space with non-commutative spacetime coordinates $\hat x^\mu$ which satisfy the commutation relations
\be
\label{deform}
\left[\hat x^\mu, \hat x^\nu \right]  =  i \Theta^{\mu\nu},
\ee
where $\Theta \equiv (\Theta^{\mu\nu})$ is a real, antisymmetric, constant matrix. In this space a general function $\hat f (\hat x) $ may be expressed using the Weyl-ordered monomials as basis elements, i.e., by the form
\be
\hat f (\hat x)= f^{(0)} + f^{(1)}_\mu :\hat x^\mu: +  f^{(2)}_{\mu\nu} :\hat x^\mu \hat x^\nu: + ...,
\ee
 where, according to the Weyl-ordering rule,
\be
\label{ordering}
:\hat x^\mu:  =  \hat x^\mu, \>\>\>\>\>\>\>\>
:\hat x^\mu \hat x^\nu:  =  \half (\hat x^\mu \hat x^\nu + \hat x^\nu \hat x^\mu),
\ee
etc. We can also introduce, in a way consistent with the relations (\ref{deform}), the operation of derivation $\hat\p_\mu \equiv \hat {\frac{\p}{\p\hat x^\mu}}$, satisfying following properties:
\bea
&&(i)~ \left[\hat \p_\mu, \hat x^\nu \right]  =   \delta^{\nu}_{\mu}, \>\>\>\>\> \>\>\>\>\> 
\left[\hat\p_\mu, \hat\p_\nu \right]  =  0 , \cr
&\cr
&&(ii)~ \hat  \p_\mu (\hat f.\hat g (\hat x))  =  (\hat \p_\mu \hat f (\hat x)) \hat g (\hat x) +  \hat f (\hat x)  \p_\mu (\hat g (\hat x)), ~~({\rm Leibniz~ rule}). 
\eea
This allows us to consider further a general higher-order differential operator $\hat X$, which acts on a function $\hat f (\hat x)$ according to
\be
\label{highdiff}
\hat X \hat f (\hat x) = \sum_r \hat\xi_r^{\mu_1 ...\mu_r} (\hat x)\hat\p_{\mu_1} ...\hat\p_{\mu_r} \hat f (\hat x)
\ee
with appropriate coefficients $\hat\xi_r^{\mu_1 ...\mu_r}(\hat x)$. 

What we have above is the algebra $\hat {\cal A}_{\hat x}$, the quotient space of the algebra freely generated by the elements $\hat x^\mu$ divided by the ideal generated by the relations (\ref{deform}). As is well-known, this algebra  $\hat {\cal A}_{\hat x}$ is isomorphic to the algebra ${\cal A}_x$ of commuting variables with the $\star$ product as multiplication. Explicitly, given two functions $\hat f (\hat x)$ and  $\hat g(\hat x)$, the isomorphism is implemented as
\be
\label{isomorphism}
\hat f (\hat x)~ \leftrightarrow ~ f (x) = f^{(0)} + f^{(1)}_\mu  x^\mu +  f^{(2)}_{\mu\nu}  x^\mu x^\nu + ... 
\ee
(similarly for $\hat g(\hat x)$), and
\be
:\hat f (\hat x) \hat g (\hat x):~ \leftrightarrow~ f (x)\star g (x) \equiv f(x) e^{\frac{i}{2}\Theta^{\mu \nu} \stackrel {\leftarrow}{\frac{\p}{\p x^\mu}}\stackrel {\rightarrow}{\frac{\p}{\p x^\nu}}} g(x).
\ee

In  \cite{Wess:2003da,Dimitrijevic:2003wv} the above isomorphism has been extended to general higher-order differential operators: there exists an invertible map from the differential operator $\hat X$ defined in $\hat {\cal A}_{\hat x}$ to a differential operator  $\Xi_X$ in ${\cal A}_x$, preserving the algebraic structure. But note that the map can change the order of the differential operators; by this reason, the strict restriction to linear differential operators (as in usual Killing symmetry considerations) becomes superfluous. Explicitly, for $\hat X \hat f$ specified as in (\ref{highdiff}), the map reads
\be
\label{mapbos}
\hat X \hat f \rightarrow X^\star \star f (x) \equiv \Xi_X f(x)
\ee
with
\be
\label{xstar}
X^\star  =  \sum_r \xi_r^{\mu_1 ...\mu_r} (x)\p_{\mu_1} ...\p_{\mu_r}, 
\ee
and
\be 
\Xi_X   =  \sum_{n,r} \left( \frac{i}{2}\right)^n \frac{1}{n!}\Theta^{\mu_1 \nu_1}...
\Theta^{\mu_n \nu_n}\xi_r^{\rho_1 ...\rho_r}(x) \left(\stackrel {\leftarrow}{\p_{\mu_1}} ... \stackrel {\leftarrow}{\p_{\mu_n}} \right) \left( \stackrel {\rightarrow} {\p_{\nu_1}} ...\stackrel {\rightarrow}{\p_{\nu_n}} \stackrel {\rightarrow}{\p_{\rho_1}} ...\stackrel {\rightarrow}{\p_{\rho_r}} \right),
\ee
assuming that $\xi_r^{\mu_1 ...\mu_r}(x)\leftrightarrow \hat\xi_r^{\mu_1 ...\mu_r}(\hat x)$ in the sense of (\ref{isomorphism}). The inverse map has also been found: if one has in the commutative space a general differential operator of the form
\be
\Xi = \sum_r \tilde\xi_r^{\mu_1 ...\mu_r} (x)\p_{\mu_1} ...\p_{\mu_r},
\ee
the $\star$ differential operator $X^\star_{\Xi}$  with the property
\be
X^\star_\Xi \star f(x) = \Xi f(x)
\ee
is given by
\be
\label{diffbos}
X^\star_\Xi = \sum_{n,r} \left(- \frac{i}{2}\right)^n \frac{1}{n!}\Theta^{\mu_1 \nu_1}...
\Theta^{\mu_n \nu_n}\tilde\xi_r^{\rho_1 ...\rho_r}(x)\left(\stackrel {\leftarrow}{\p_{\mu_1}} ... \stackrel {\leftarrow}{\p_{\mu_n}} \right) \left(\stackrel {\rightarrow} {\p_{\nu_1}} ...\stackrel {\rightarrow}{\p_{\nu_n}} \stackrel {\rightarrow}{\p_{\rho_1}} ...\stackrel {\rightarrow}{\p_{\rho_r}} \right).
\ee
From $X^\star_\Xi$ follows $\hat X_\Xi$ immediately (i.e., write the expression (\ref{diffbos}) in the form (\ref{xstar}) and then use above the correspondence to change it into the corresponding noncommutative-space differential operator $\hat X$). Note that if $\Xi$ is a linear differential operator but with coefficients which are lth-degree polynomials in $x$, we will end up with the operator $\hat X_\Xi$ which is (l+1)th-order differential operator.

 The inverse map given above can be used to construct the differential operator representations (or space-time representations) for the symmetry algebra in $\hat {\cal A}_{\hat x}$. Take the case of conformal generators, encompassing those for the translations (P), Lorentz transformations (M), the dilatation (D) and special conformal transformations (K). In the commutative space (i.e. with $\Theta = 0$), they are represented by the linear differential operators:
\bea
\label{dddd}
 P_\mu & = & - i  \p_\mu, \>\>\>\>\>\>\>
 M_{\mu\nu}  =  - i ( x_\mu \p_\nu -  x_\nu  \p_\mu), \cr 
&\cr
D & = & i  x^\mu  \p_\mu, \>\>\>\>\>\>\> K_\mu  =  - i(2  x_\mu  x^{\nu}  - \delta^\nu_\mu  x^2 ) \p_\nu .
\eea
Identifying $\Xi$ with any of these operators, we can now apply the inverse map to obtain the related generators in $\hat {\cal A}_{\hat x}$, viz., 
\bea
\label{deformcon}
\hat P_\mu & = & - i \hat \p_\mu,\>\>\>\>\>\>\>\>\>\ \hat D =  i \hat x^\mu \hat \p_\mu, \cr
&\cr
\hat M_{\mu\nu} & = & - i (\hat x_\mu \hat\p_\nu - \hat x_\nu \hat \p_\mu) 
 -  \half (\Theta^{\rho\lambda}\eta_{\rho\mu}\hat \p_\nu \hat \p_\lambda - \Theta^{\rho\lambda}\eta_{\rho\nu}\hat\p_\mu \hat\p_\lambda), \cr 
&\cr
\hat K_\mu & = & - 2 i \hat x_\mu \hat x^{\nu}\hat \p_\nu + i \hat x^2 \hat \p_\mu - \Theta^{\rho\lambda}\eta_{\rho\mu} \hat \p_\lambda  - \Theta^{\rho\lambda}\eta_{\rho\mu} \hat x^\nu \hat \p_\nu \hat \p_\lambda \cr
&\cr
&& +  \Theta^{\rho\lambda} \hat x_\rho \hat \p_\lambda \hat\p_\mu - \frac{i}{4}\eta_{\rho\lambda} \Theta^{\rho\sigma} \Theta^{\lambda\tau}\hat\p_\sigma\hat\p_\tau\hat\p_\mu.\eea
Notice that $\hat M_{\mu\nu}$ contains double derivatives and $\hat K_\mu$ terms up to triple derivatives; these follow since the corresponding undeformed forms in (\ref{dddd}) involve coefficients which are linear and quadratic in $x$, respectively. These deformed generators satisfy the usual commutation relations of the conformal algebra (which we verified explicitly):
\bea
\label{boscon}
\left[\hat \p_\mu, \hat \p_\nu \right] & = & 0 , \>\>\>\>\> \left[\hat \delta_\omega, \hat \p_\nu \right]  =   \omega^{~\lambda}_\nu\hat\p_\lambda, \cr
&\cr
\left[\hat \delta_\omega, \hat \delta^\prime_\omega \right] & =&  \hat\delta_{\omega\times\omega^\prime},~~ ({\rm with}~ \omega\times\omega^\prime \equiv -(\omega^{~\sigma}_\mu\omega^{\prime\nu}_\sigma - \omega^{\prime\sigma}_\mu\omega^{~\nu}_\sigma)) \cr
&\cr
\left[\hat \delta_d, \hat \p_\nu \right] & = &  \hat \p_\nu, \>\>\>\>\> \left[\hat \delta_d, \hat \delta_c \right]  =  - \hat \delta_c, \cr&\cr
\left[\hat \p_\mu, \hat \delta_c \right] & = & - 2  c_\mu\hat \delta_d + 2 i c^\nu  \hat M_{\nu\mu},\cr
&\cr
\left[\hat \delta_\omega, \hat \delta_c \right] & = &  i c^\nu\omega^{~\lambda}_\nu\hat K_\lambda, \cr
&\cr
\left[\hat \delta_d, \hat \delta_\omega \right] & = & \left[\hat \delta_{c}, \hat \delta^\prime_{c} \right] = 0,
\eea
where $\hat\delta_d, \hat\delta_\omega$ and $\hat\delta_c$ denote the infinitesimal generators defined as ($\omega_{\mu\nu}, c_\lambda $ are real constants): 
\be
\label{deformconrep}
\hat \delta_d ~= i \hat D, ~~~~ \hat\delta_\omega  = -\frac{i}{2}\omega_{\mu\nu}\hat M^{\mu\nu}, ~~~
\hat\delta_c = i c_\lambda\hat K^{\lambda} 
\ee
We remark that if, instead of the operators given in (\ref{deformcon}), the related higher-derivative $\star$ differential operators $(\p^\star_\mu, \delta^\star_d, \delta^\star_\omega, \delta^\star_c)$ are considered,they also satisfy the unmodified conformal algebra relations (i.e., in the form (\ref{boscon})) with the bracket now provided by the so-called $\star-$commutator.
 
There is an alternative way to understand the higher derivative structure that we have found for the generators in the deformed space. In the abstract phase space having $(x^\mu,p_\mu)$ as canonical coordinates, it is possible to introduce {\it noncanonical} coordinates $(\hat x^\mu,\hat p_\mu)$ which are related to $( x^\mu, p_\mu)$ by
\begin{eqnarray}
\label{Darb}
\hat{x}^{\mu} &=& x^{\mu} - \frac{1}{2}\Theta^{\mu \sigma}p_{\sigma}, \cr
&\cr
\hat{p}_{\mu} &=& p_{\mu}.
\end{eqnarray}
Then the Poisson brackets between the $x's$ and $p's$ will be non-standard ones -- we rather find, after quantization, the  ones consistent with the noncommutative-space commutation relations (\ref{deform}), together with $\left[\hat p_\mu, \hat p_\nu \right] = 0$ and  $\left[\hat p_\mu, \hat x^\nu \right] = - i \delta^\nu_\mu$. The inverse transformation of (\ref{Darb}) reads
\begin{eqnarray}
\label{Darb1}
x^{\mu} &=& \hat{x}^{\mu} + \frac{1}{2}\Theta^{\mu \sigma}\hat{p}_{\sigma},
\cr
&\cr
p_{\mu} &=& \hat{p}_{\mu} = -i \hat \p_\mu ,
\end{eqnarray}
where we have made the replacements $\hat{p}_{\mu} \rightarrow -i \hat \p_\mu$. If we use the transformation (\ref{Darb1}) in (\ref{dddd}), we then obtain, quite remarkably, the deformed generators shown in (\ref{deformcon}); higher derivative terms show up here because our transformation  (\ref{Darb1}) mixes coordinates with momenta. This also explains why our deformed generators satisfy the unmodified commutation relations of the conformal algebra.

To make connection with the Hopf-algebra based approach \cite{Chaichian:2004za}, we now derive the comultiplication rules implied by the above differential representation. They follow upon applying the operators in (\ref{deformcon}) on the product of two functions, $\hat f \hat g$. Here, for the coproduct of translation generators $\Delta(\hat P_\mu)$, the Leibniz rule holds (since $\hat P_\mu$ equals $(-i\hat\p_\mu)$:
\be
\label{cp0}
\Delta(\hat P_\mu) = \hat P_\mu \otimes {\bf{1}} + {\bf{1}} \otimes \hat P_\mu.
\ee
On the other hand, using the above expressions for $\hat M_{\mu\nu}$ and $\hat D$, which involve higher derivative operators, we find
\bea
\hat M_{\mu\nu} (\hat f \hat g) & = & (\hat M_{\mu\nu}\hat f)\hat g + \hat f(\hat M_{\mu\nu}\hat g) -  \frac{1}{2}\left(\Theta^{~\sigma}_\mu \hat \p_\nu -\Theta^{~\sigma}_\nu \hat \p_\mu \right) \hat f \hat\p_\sigma \hat g \cr
&\cr
& + & \frac{1}{2} \hat \p_\sigma \hat f \left(\Theta^{~\sigma}_\mu \hat \p_\nu -\Theta^{~\sigma}_\nu \hat \p_\mu \right) \hat g , \cr
&\cr
\hat D(\hat f \hat g) & = & (\hat D\hat f)\hat g + \hat f(\hat D\hat g)
+\Theta^{\mu\nu}\hat\partial_\nu\hat f
\hat\partial_\mu\hat g ,
\label{cp}
\eea
and these in turn lead to the following comultiplication rules:
\bea
\Delta(\hat M_{\mu\nu})& = & \hat M_{\mu\nu} \otimes {\bf{1}} + {\bf{1}} \otimes \hat M_{\mu\nu} +  \frac{1}{2}\left(\Theta^{~\sigma}_\mu \hat P_\nu -\Theta^{~\sigma}_\nu \hat P_\mu \right)\otimes \hat P_\sigma \cr
&\cr
& - & \frac{1}{2} \hat P_\sigma  \otimes \left(\Theta^{~\sigma}_\mu \hat P_\nu -\Theta^{~\sigma}_\nu \hat P_\mu \right) ,  \cr
&\cr
\Delta(\hat D) & = &\hat D \otimes {\bf{1}} + {\bf{1}} \otimes \hat D
+ \Theta^{\mu\nu}\hat P_\mu\otimes \hat P_\nu .
\label{cp1}
\eea
Similarly the comultiplication rules for the special conformal generators $\hat K_{\mu}$
are obtained as
\begin{eqnarray}
\Delta(\hat K_\mu) & = & \hat K_\mu \otimes {\bf{1}} + {\bf{1}} \otimes \hat K_\mu \cr
&\cr
&+& \Theta^{\rho\sigma}\Big((\eta_{\rho\mu}\hat D +\hat M_{\rho\mu})\otimes \hat P_\sigma +  \hat P_\rho\otimes (\eta_{\sigma\mu}\hat D +\hat M_{\sigma\mu})\Big) \cr&\cr
& + & \frac{1}{4}\Theta^{\rho\sigma}\Theta^{\lambda\omega}\Big((\eta_{\lambda\mu}\hat P_\rho
-\eta_{\lambda\rho} \hat P_\mu +\eta_{\rho\mu} \hat P_\lambda)\otimes \hat P_\omega \hat P_\sigma \cr
&\cr
&& +  \hat P_\lambda \hat P_\rho \otimes (\eta_{\omega\mu}\hat P_\sigma
-\eta_{\omega\sigma} \hat P_\mu +\eta_{\sigma\mu} \hat P_\omega)\Big) .
\label{coproduct2}
\end{eqnarray}
In this manner, i.e., with the help of explicit differential realizations, the comultiplication rule $\Delta(\hat M_{\mu\nu})$ (first appeared in \cite{Chaichian:2004za} ) was deduced in \cite{Wess:2003da}  (see also \cite{Koch:2004ud}). As for the rules involving $\hat D$ and $\hat K_{\mu}$, the results given above (which we derived using our differential operator realizations in (\ref{deformcon})) were also found recently \cite{Matlock:2005zn} from quantum group arguments. One can easily verify that the coproducts $ (\Delta(\hat P_\mu),\Delta(\hat M_{\mu\nu},\Delta(\hat D),\Delta(\hat K_\mu))$, with comultiplication rules given above, satisfy the commutation relations of the conformal algebra without modification $-$ they define the conformal bialgebra.

\section{Deformed superspace calculus and super-Poincar\'e and superconformal algebras}

Let us now turn to the case of non-(anti)commutative superspace. In this work we shall be concerned with the unextended (i.e. ${\mathcal{N}} = 1$) superspace only. Before considering any deformation, the coordinates of this superspace are denoted by $(x^\mu, \theta^\alpha, \bar\theta^{\dot\alpha})$. (We follow the notation of \cite{wess}). The Grassmannian coordinates $\theta^\alpha$ and $\bar\theta^{\dot\alpha}$ anticommute; but in the deformed superspace, we may assume the related coordinates $\hat\theta^\alpha,\hat{\bar\theta^{\dot\alpha}}$ to satisfy the anticommutation relations of the form \cite{Ferrara:2000mm,Seiberg:2003yz}
\be 
\label{deformferm}
\left\{\hat\theta^\alpha, \hat\theta^\beta \right\}  =   C^{\alpha\beta}, \>\>\>\>\>\>
\left\{\hat{\theta}^{\alpha}, \hat{\bar\theta}^{\dot\alpha} \right\}= \left\{\hat{\bar\theta}^{\dot\alpha}, \hat{\bar\theta}^{\dot\beta} \right\}  =   0.
\ee
with nonzero constant $C^{\alpha\beta}$. This deformation, breaking half of ${\mathcal{N}} = 1$ supersymmetry, will lead to a theory with  ${\mathcal{N}} = \frac{1}{2}$ supersymmetry \cite{Seiberg:2003yz}.

To discuss more general deformation (containing the deformed commutation relations (\ref{deform})), 
it is convenient to introduce the chiral coordinates
\be
\hat y^\mu = \hat x^\mu + i \hat\theta^\alpha \sigma^\mu_{\alpha \dot{\alpha}} \hat{\bar{\theta}}^{\dot{\alpha}}
\ee
and then posit the (anti-)commutation relations \cite{Ferrara:2000mm,Seiberg:2003yz}
\be
\label{deformferm1}
\left[\hat y^\mu, \hat y^\nu \right]  =  i \Theta^{\mu\nu}, ~~~~~
\left[\hat y^\mu, \hat\theta^\alpha \right]  =  \Psi^{\mu\alpha}, ~~~~~
\left[\hat y^\mu, \hat{\bar\theta}^{\dot\alpha} \right]  =  0
\ee
together with (\ref{deformferm}). Here, $\Psi^{\mu\alpha}$ are some Grassmannian c-numbers. The result is the deformed superspace with three kind of deformation parameters, i.e., $\Theta^{\mu\nu}, C^{\alpha\beta}$ and $\Psi^{\mu\alpha} $.

To discuss functions on the deformed superspace, we need to introduce appropriate ordering on related monomials. Here we adopt the natural generalization of (\ref{ordering}), i.e.,
\bea
\label{orderingferm}
:\hat y^\mu \hat y^\nu:  & = &  \half (\hat y^\mu \hat y^\nu + \hat y^\nu \hat y^\mu), \cr
&\cr
:\hat \theta^\alpha \hat \theta^\beta:  & = &  \half (\hat\theta^\alpha\hat \theta^\beta - \hat \theta^\beta\hat\theta^\alpha), \cr
&\cr
:\hat y^\mu \hat \theta^\alpha: & = & \half (\hat y^\mu \hat\theta^\alpha + \hat\theta^\alpha \hat y^\mu) ,
\eea
etc. Since the status of the coordinates $\hat{\bar\theta}^{\dot\alpha}$ is not changed in our deformation, no ordering rules for them need to be specified. In this space, we can again introduce (in a way consistent with the relations (\ref{deformferm}) and (\ref{deformferm1})) the derivatives $\hat\p_\mu \equiv \frac{\p}{\p \hat y^\mu}|_{\hat\theta,\hat{\bar\theta}}, \hat\p_\alpha\equiv \frac{\p}{\p \hat\theta^\alpha}|_{\hat y,\hat{\bar\theta}}$ and ${\hat{\bar\p}}_{\dot\alpha}\equiv \frac{\p}{\p \hat{\bar{\theta}}^{\dot\alpha}}|_{\hat y,\hat \theta}$ satisfying
\be
\left[\hat\p_\mu, \hat y^\nu \right]   =  \delta^{\nu}_{\mu}, \>\>\>\>\>  \left\{\hat \p_\alpha, \hat \theta^\beta \right\}  =  \delta^{\beta}_{\alpha},\>\>\>\>\> \left\{\hat\p_{\dot\alpha}, \hat{\bar{\theta}}^{\dot\beta} \right\}  =  \delta^{\dot\beta}_{\dot\alpha} ,
\ee
plus other vanishing (anti)commutation relations. Notice that $\hat\p_\alpha$ and $\hat{\bar\p}_{\dot\alpha}$ denote partial derivatives at 'fixed' $\hat y^\mu$ rather than at 'fixed' $\hat x^\mu$. Given due consideration on the Grassmann parity nature of the functions involved, the Leibniz rule holds with these derivatives.

We now discuss the isomorphism existing between the algebra of functions in the deformed superspace and that defined in the ordinary superspace. Here, let us denote the deformed superspace coordinates by $\hat Y^M \equiv (\hat y^\mu, \hat \theta^\alpha, \hat{\bar\theta}^{\dot\alpha})$, and the corresponding undeformed ones by $Y^M \equiv (y^\mu, \theta^\alpha, \bar\theta^{\dot\alpha})$. For the ordered product considered in (\ref{orderingferm}), we write $:\hat Y^M \hat Y^N:$. Then a general function of the deformed superspace (with definite Grassmann parity), $\hat F (\hat Y)$, may be expressed by the form (here $:\hat Y^M: \equiv \hat Y^M$)
\be
\hat F (\hat Y) =  f^{(0)} + f^{(1)}_M :\hat Y^M: +  f^{(2)}_{MN} :\hat Y^M \hat Y^N: + ...~ .
\ee
Now the isomorphism is implemented by
\be
\label{iso}
\hat F (\hat Y) ~ \leftrightarrow ~ F (Y) = f^{(0)} + f^{(1)}_M  Y^M +  f^{(2)}_{MN}  Y^M Y^N + ...~ ,
\ee
and, for the product $:\hat F (\hat Y) \hat G ( \hat Y):$ involving two arbitrary functions $\hat F (\hat Y)$ and $ \hat G ( \hat Y)$, by the generalized $\star$ product
\be
:\hat F (\hat Y)\hat G (\hat Y): ~\leftrightarrow ~ F\star G(Y) \equiv F(Y) e^{\frac{i}{2}\Theta^{M N}\stackrel {\leftarrow}{\frac{\p}{\p Y^M}} \stackrel {\rightarrow}{\frac{\p}{\p Y^N}}} G(Z)|_{Y\rightarrow Z}~ ,
\ee
where $\stackrel {\rightarrow} {\frac{\p}{\p Y^N}} \equiv (\p_\mu \equiv \frac{\p}{\p y^\mu}|_{\theta,\bar\theta}, \p_\alpha\equiv \frac{\p}{\p \theta^\alpha}|_{y,\bar\theta},{\bar\p}_{\dot\alpha}\equiv \frac{\p}{\p \bar{\theta}^{\dot\alpha}}|_{y,\theta}), \stackrel {\leftarrow} {\frac{\p}{\p Y^N}} \equiv (\stackrel {\leftarrow}{\p_\mu},\stackrel {\leftarrow}{\p_\alpha}, \stackrel {\leftarrow}{\p_{\dot\alpha}}) $ with ($\theta^\alpha\stackrel{\leftarrow}{\p_\beta} = - \delta^\alpha_\beta$, etc.) and $\Theta \equiv (\Theta^{M N})$ is the supermatrix
\be
\left(\begin{array}{ccc} \Theta^{\mu\nu} & -i \Psi^{\mu\beta} & 0 \\ i \Psi^{\alpha\nu} & i C^{\alpha\beta} & 0 \\ 0 & 0 & 0 \end{array}\right)
\ee
with $\Psi^{\alpha\nu} = - \Psi^{\nu\alpha}$. Similarly, an invertible map preserving the algebraic structure can be established between a differential operator $\hat {X}$ defined in the deformed superspace and a related differential operator $\Xi_{X}$ in the ordinary superspace. Let $\hat {X}$ be a general differential operator (with definite Grassman parity), which acts on a function $\hat F (\hat Y)$ as
\be
\hat {X} \hat F (\hat Y) = \sum_r \hat\xi_r^{M_1 ...M_r} (\hat Y)\hat\p_{M_1} ...\hat\p_{M_r} \hat F(Y)
\ee
with appropriate coefficients $\hat\xi_r^{M_1 ...M_r}(\hat Y)$. Here we have denoted $\hat\p_M \equiv (\hat\p_\mu, \hat\p_\alpha, \hat{\bar\p}_{\dot\alpha})$. This is then mapped to the elements $\Xi_X$ of the ordinary superspace by
\be
\label{mapferm}
\hat {X} \hat F(\hat Y)  \rightarrow  {X^\star} \star F (Y) \equiv \Xi_{X} F (Y),
\ee
where, if $\xi_r^{M_1 ...M_r}(Y)$ is related to  $\hat \xi_r^{M_1 ...M_r}(\hat Y)$ in the sense of (\ref{iso}), we have
\be
{X^\star} = \sum_r \xi_r^{M_1 ...M_r}  (Y)\p_{M_1} ...\p_{M_r}
\ee
and therefore
\be
\Xi_{X} = \sum_{n,r} \left( \frac{i}{2}\right)^n \frac{1}{n!}\Theta^{M_1 N_1}...
\Theta^{M_n N_n}\xi_r^{\rho_1 ...\rho_r}  (Y)\left(\stackrel {\leftarrow}{\p_{M_1}} ... \stackrel {\leftarrow}{\p_{M_n}} \right) \left( \stackrel {\rightarrow} {\p_{N_1}} ...\stackrel {\rightarrow}{\p_{N_n}} \stackrel {\rightarrow}{\p_{\rho_1}} ...\stackrel {\rightarrow}{\p_{\rho_r}} \right).
\ee
The map (\ref{mapferm}), an exact parallel of the result (\ref{mapbos}) without Grassmannian coordinates, can be demonstrated, say, by the method of induction \cite{Wess:2003da}. For the inverse map, given a differential operator
\be
\label{diffferm}
\Xi  = \sum_r \xi_r^{M_1 ...M_r}  (Y)\p_{M_1} ...\p_{M_r} ,
\ee
we must find a differential operator $X^\star_\Xi$ such that
\be
X^\star_\Xi \star F (Y) = \Xi F (Y).
 \ee
The desired differential operator can be shown to be of the form (\ref{diffbos}), i.e.
\be
\label{diffferm1}
X^\star_\Xi = \sum_{n,r} \left(- \frac{i}{2}\right)^n \frac{1}{n!}\Theta^{M_1 N_1}...
\Theta^{M_n N_n} \xi_r^{\rho_1 ...\rho_r}(Y)\left(\stackrel {\leftarrow}{\p_{M_1}} ... \stackrel {\leftarrow}{\p_{M_n}} \right) \left( \stackrel {\rightarrow} {\p_{N_1}} ...\stackrel {\rightarrow}{\p_{N_n}} \stackrel {\rightarrow}{\p_{\rho_1}} ...\stackrel {\rightarrow}{\p_{\rho_r}} \right) ,
\ee
as may be verified by checking whether $\Xi_{X}$ for ${X}$ given by the expression (\ref{diffferm1}) leads back to the form (\ref{diffferm}). Promoting $X^\star_\Xi$ to $\hat X_\Xi$ is trivial.

We are now ready to give the differential operator representation for the super-Poincar\'e and superconformal generators in the superspace. As for the generators of the super-Poincar\'e algebra, we may apply the above inverse map to the corresponding expressions in the ordinary superspace, i.e., to
\bea
\label{Poincare0}
P_\mu & = & - i \p_\mu, \>\>\>\>\>\> Q_\alpha  =  i \p_\alpha, \cr
&\cr
{\bar{Q}}_{\dot{\alpha}} & = &  - i {\bar{\p}}_{\dot{\alpha}} -2 \theta^\alpha \sigma^\rho_{\alpha \dot{\alpha}} \p_\rho , \cr 
&\cr
M_{\mu\nu} & = & - i ( y_\mu \p_\nu - y_\nu \p_\mu) +  i \theta^\alpha (\sigma_{\mu\nu})_{\alpha}^{~\beta} \p_{\beta} - i  {\bar{\theta}^{\dot{\alpha}}} (\bar{\sigma}_{\mu\nu})^{\dot{\beta}}_{~\dot{\alpha}}{\bar{\p}}_{\dot{\beta}}.
\eea
The result is (here $\hat\p_{\alpha\dot\alpha} \equiv \sigma^\rho_{\alpha\dot\alpha}\hat\p_\rho$)
\bea
\label{Poincare}
\hat P_\mu & = & - i \hat \p_\mu, \cr 
&\cr
\hat M_{\mu\nu} & = & - i (\hat y_\mu \hat\p_\nu - \hat y_\nu \hat\p_\mu) +  i \hat\theta^\alpha (\sigma_{\mu\nu})_{\alpha}^{~\beta} \hat\p_{\beta} -  i  \hat{\bar{\theta}}^{\dot{\alpha}} (\bar{\sigma}_{\mu\nu})^{\dot{\beta}}_{~\dot{\alpha}}\hat{\bar{\p}}_{\dot{\beta}} \cr
&\cr
& - & \half (\Theta^{~\rho}_\mu\hat\p_\nu \hat\p_\rho - \Theta^{~\rho}_\nu\hat\p_\mu \hat\p_\rho) + \frac{i}{2} (\Psi^{~\alpha}_\mu\hat\p_\nu \hat\p_\alpha - \Psi^{~\alpha}_\nu\hat\p_\mu \hat\p_\alpha) \cr&\cr
& + &  \frac{i}{2} \Psi^{\rho\alpha}(\sigma_{\mu\nu})_{\alpha}^{~\beta} \hat\p_\rho \hat\p_\beta -  \frac{i}{2} C^{\alpha\beta}(\sigma_{\mu\nu})_{\alpha}^{~\gamma} \hat\p_\beta \hat\p_\gamma ,\cr 
&\cr
\hat Q_\alpha & = & i \hat\p_\alpha ,
\cr
&\cr
\hat{\bar{Q}}_{\dot{\alpha}}  & = &  - i \hat{\bar{\p}}_{\dot{\alpha}} -2 \hat\theta^\alpha \hat\p_{\alpha \dot{\alpha}} + C^{\alpha\beta}\hat\p_\beta \hat\p_{\alpha\dot{\alpha}} - \Psi^{\rho\alpha}\hat\p_\rho \hat\p_{\alpha\dot{\alpha}}.
\eea
As one can verify explicitly, this differential operator representation in the deformed superspace satisfy the unmodified (anti)commutation relations of the super-Poincar\'e algebra
\bea
\label{fermPoin}
\left[\hat P_\mu, \hat Q_\alpha \right] & = & \left[\hat P_\mu, \hat {\bar{Q}}_{\dot{\alpha}} \right]  =  0 ,\cr
&\cr
\left[\hat M_{\mu\nu}, \hat Q_{\alpha} \right] & = & - i (\sigma_{\mu\nu})_{\alpha}^{~\beta} \hat Q_{\beta} ,\cr
&\cr
\left[\hat M_{\mu\nu}, \hat {\bar{Q}}_{\dot{\alpha}} \right] & = &  i (\bar{\sigma}_{\mu\nu})_{~\dot{\alpha}}^{\dot{\beta}} \hat{\bar{Q}}_{\dot{\beta}} ,\cr
&\cr
\left\{\hat Q_{\alpha}, \hat {\bar{Q}}_{\dot{\alpha}} \right\} & = & 2 \hat P_{\alpha\dot{\alpha}} .
\eea

Are there appropriate analogues of transformations (\ref{Darb}) and (\ref{Darb1}) in the superspace ? The answer is yes. From (\ref{deformferm}) and (\ref{deformferm1}) it is clear that only ${y}^\mu$ and ${\theta^{\alpha}}$ need to be transformed while other variables (including derivatives) remain unchanged. We then find, for the relevant transformations, following formulas:
\begin{eqnarray}\hat{y}^{\mu} &=& y^{\mu} + \frac{i}{2} \Theta^{\mu \nu} \partial_{\nu} + \frac{1}{2} \Psi^{\mu \alpha} \partial_{\alpha},
\cr
&\cr
\hat{\theta}^{\alpha} &=& \theta^{\alpha} + \frac{1}{2} C^{\alpha \beta} \partial_{\beta} + \frac{1}{2} \Psi^{\alpha\mu} \partial_{\mu}.
\end{eqnarray}
Taking the usual canonical brackets for $y$, $\theta$, etc. immediately reproduces the noncanonical (deformed) algebra of (\ref{deformferm}) and (\ref{deformferm1}). Since the derivatives are invariant, the inverse transformations are given by
\begin{eqnarray}
y^{\mu} &=& \hat{y}^{\mu} - \frac{i}{2} \Theta^{\mu \nu} \hat{\partial}_{\nu} - \frac{1}{2} \Psi^{\mu \alpha} \hat{\partial}_{\alpha}, \cr
&\cr
\theta^{\alpha} &=& \hat{\theta}^{\alpha} - \frac{1}{2} C^{\alpha \beta} \hat{\partial}_{\beta} - \frac{1}{2} \Psi^{\alpha\mu} \hat{\partial}_{\mu}.
\end{eqnarray}
Substituting these expressions in (\ref{Poincare0}) leads to the deformed generators (\ref{Poincare}). It also explains the fact that these deformed generators satisfy the same super-Poincar\'e algebra as the undeformed ones. Hence, despite deformation, one may use the same Casimir operators as the undeformed case to classify various representations, etc.

We can apply the inverse map to the superconformal generators in the ordinary superspace, to get a representation of the superconformal algebra in the deformed superspace. Appropriate linear differential operator representation in the ordinary superspace can be found for instance in \cite{buchbinder}. Because of the complexity involved in the expressions, we shall present the representation with the $C^{\alpha\beta}$ deformation only. (If one wishes, the full expressions containing deformation parameters $\Theta^{\mu\nu}$ and $\Psi^{\mu\alpha}$ as well can be easily worked out). The superconformal algebra has 24 generators -- in addition to 14 generators $(P_\mu,M_{\mu\nu},Q_{\alpha},{\bar{Q}}_{\dot{\alpha}})$ of the super-Poincar\'e algebra, it contains $D$ (dilatation), $K_\mu$ (special conformal transformations), $A$ (axial charge) and $S_\alpha ,{\bar S}_{\dot\alpha}$ (so-called S-supersymmetry transformations). In the C-deformed superspace the representations of $(P_\mu,M_{\mu\nu},Q_{\alpha},{\bar{Q}}_{\dot{\alpha}})$ are available from (\ref{Poincare}), now with $\Theta^{\mu\nu}= \psi^{\mu\alpha} = 0$. For the other generators the inverse map gives rise to the following representations:
\bea
\label{deformscon}
\hat D & = & i \hat y^\mu \hat\p_\mu + \frac{i}{2} \hat\theta^\alpha \hat\p_\alpha + \frac{i}{2} \hat{\bar{\theta}}^{\dot{\alpha}} \hat{\bar{\p}}_{\dot{\alpha}} , \>\>\>\>\>\> \hat A  =  \frac{1}{2} \left(\hat\theta^\alpha \hat\p_\alpha - \hat{\bar{\theta}}^{\dot{\alpha}} \hat{\bar{\p}}_{\dot{\alpha}} \right) ,\cr 
&\cr
\hat K_{\alpha\dot\alpha} & = & - i (\hat y)^{~\dot\beta}_\alpha (\hat y)^{\beta}_{~\dot\alpha} \hat \p_{\beta\dot\beta} 
 + 2 i \theta_\alpha (\hat y)^{\beta}_{~\dot\alpha} \hat \p_\beta + 2 i \hat{\bar\theta}_{\dot\alpha} (\hat y)^{~\dot\beta}_\alpha \hat{\bar\p}_{\dot\beta} 
 +  4  \hat\theta_\alpha \hat{\bar\theta}^2 \hat{\bar\p}_{\dot\alpha} \cr
&\cr
& - & i \epsilon_{\alpha\gamma} C^{\gamma\delta} (\hat y)^{\beta}_{~\dot\alpha}\hat \p_\delta\hat\p_\beta - 2  \epsilon_{\alpha\gamma} C^{\gamma\delta}\hat{\bar\theta}^2 \hat\p_\delta\hat{\bar\p}_{\dot\alpha} ,\cr 
&\cr
\hat S_\alpha & = & -2 i (\hat y)^{~\dot\beta}_\alpha \hat \theta^\beta \hat \p_{\beta\dot\beta} + 2 i \hat \theta^2 \hat\p_\alpha + (\hat y)^{~\dot\beta}_\alpha \hat {\bar{\p}}_{\dot\beta} + 4 i \hat \theta_\alpha \hat{\bar\theta}^{\dot\beta}\hat{\bar{\p}}_{\dot\beta} \cr
&\cr  
& + &  i C^{\beta\gamma}(\hat y)^{~\dot\beta}_\alpha \hat \p_\gamma  \hat \p_{\beta\dot\beta} + 2 i C^{\beta\gamma} \hat \theta_\beta\hat\p_\gamma\hat\p_\alpha 
 -  2 i \epsilon_{\alpha\beta}  C^{\beta\gamma} \hat{\bar\theta}^{\dot\beta}\hat{\bar\p}_{\dot\beta}\hat\p_\gamma ,\cr
&\cr
\hat{\bar S}_{\dot\alpha} & = & - 2 i \hat{\bar\theta}^2 \hat{\bar\p}_{\dot\alpha} + (\hat y)^{\beta}_{~\dot\alpha} \hat\p_\beta ,
\eea
where $\hat y_\alpha^{~\dot\beta} = \hat y^\mu (\sigma_\mu)_\alpha^{~\dot\beta}$, etc. The (anti)commutation relations of the superconformal algebra are satisfied by this representation without modification; viz., they satisfy (in addition to those in (\ref{fermPoin}) )
\bea
\label{deformscon1}
\left[\hat D, \hat Q_\alpha \right] & = & -\frac{i}{2} \hat Q_\alpha, ~~~~~~~\left[\hat D, \hat {\bar{Q}}_{\dot{\alpha}} \right]  =  -\frac{i}{2} \hat{\bar{Q}}_{\dot{\alpha}}, \cr
&\cr
\left[\hat D, \hat S_\alpha \right] & = & \frac{i}{2} \hat S_\alpha, ~~~~~~~~~~
\left[\hat D, \hat {\bar{S}}_{\dot{\alpha}} \right]  =  \frac{i}{2} \hat{\bar{S}}_{\dot{\alpha}}, \cr
&\cr
\left[\hat P_{\alpha\dot\alpha},\hat S_\beta \right] & = & 2\epsilon_{\alpha\beta}\hat{\bar{Q}}_{\dot{\alpha}},~~~~~~
\left[\hat P_{\alpha\dot\alpha},\hat {\bar{S}}_{\dot{\beta}} \right]  =  - 2\epsilon_{\dot\alpha\dot\beta} \hat Q_\alpha, \cr
&\cr
\left[\hat K_{\alpha\dot\alpha}, \hat Q_\beta \right] & = & 2\epsilon_{\alpha\beta}\hat {\bar{S}}_{\dot{\alpha}},~ ~~~~~~
\left[\hat K_{\alpha\dot\alpha},\hat {\bar{Q}}_{\dot{\beta}} \right]  =  - 2\epsilon_{\dot\alpha\dot\beta} \hat S_\alpha, \cr
&\cr
\left [\hat M_{\mu\nu}, \hat S_{\alpha} \right] & = &  i (\sigma_{\nu\nu})_{\alpha}^{~\beta} \hat Q_{\beta},  ~~~
\left[\hat M_{\mu\nu}, \hat {\bar{S}}_{\dot{\alpha}} \right]  =   i (\bar{\sigma}_{\nu\nu})_{~\dot{\alpha}}^{\dot{\beta}} \hat {\bar{Q}}_{\dot{\beta}}, \cr
&\cr
\left\{\hat S_{\alpha}, \hat{\bar{S}}_{\dot{\alpha}} \right\} & = & 2 \hat K_{\alpha\dot\alpha}, \cr
&\cr
\left\{\hat S_{\alpha}, \hat Q_\beta \right\}  & = &  4 i \hat M_{\alpha\beta} + 2 i \epsilon_{\alpha\beta}\hat D + 6 \epsilon_{\alpha\beta}\hat A .
\eea

While the (anti)commutation relations of the symmetry algebra are unchanged, there are modifications to the comultiplication structure because of the deformation. The comultiplication rules for our super-Poincar\'e generators follow if we apply the differential operators in (\ref{Poincare}) on the product of, say, two scalar superfields: the results are
\bea
\label{comsusy}
\Delta(\hat P_{\mu})& = & \hat P_{\mu} \otimes {\bf{1}} + {\bf{1}} \otimes \hat P_{\mu} ,\cr
&\cr
\Delta(\hat M_{\mu\nu})& = & \hat M_{\mu\nu} \otimes {\bf{1}} + {\bf{1}} \otimes \hat M_{\mu\nu} \cr
&\cr
& + & \frac{1}{2}\left(\Theta^{~\sigma}_\mu \hat P_\nu -\Theta^{~\sigma}_\nu \hat P_\mu \right)\otimes \hat P_\sigma 
 -  \frac{1}{2} \hat P_\sigma \otimes \left(\Theta^{~\sigma}_\mu \hat P_\nu -\Theta^{~\sigma}_\nu \hat P_\mu \right) \cr
&\cr
& + & \frac{i}{2}\left( \Psi^{~\alpha}_\mu P_\nu - \Psi^{~\alpha}_\nu P_\mu \right)\otimes Q_\alpha - \frac{i}{2} Q_\alpha \otimes \left( \Psi^{~\alpha}_\mu P_\nu - \Psi^{~\alpha}_\nu P_\mu \right) \cr
\cr
&-& \frac{i}{2}\Psi^{\rho\alpha}(\sigma_{\mu\nu})_{\alpha}^{~\beta} \left(P_\rho \otimes Q_\beta - Q_\beta  \otimes P_\rho \right) \cr
&\cr
& - & \frac{i}{2} C^{\alpha\beta}(\sigma_{\mu\nu})_{\alpha}^{~\gamma}\left(Q_\beta \otimes Q_\gamma + Q_\gamma\otimes Q_\beta \right) ,\cr  
&\cr
\Delta(\hat{{Q}}_{{\alpha}}) & = & \hat{{Q}}_{{\alpha}}\otimes{\bf{1}} + {\bf{1}}\otimes\hat{{Q}}_{{\alpha}} ,\cr
&\cr
\Delta(\hat{\bar{Q}}_{\dot{\alpha}}) & = & \hat{\bar{Q}}_{\dot{\alpha}}\otimes{\bf{1}} + {\bf{1}}\otimes\hat{\bar{Q}}_{\dot{\alpha}} + C^{\alpha\beta}\left(\sigma^\rho_{\alpha\dot{\alpha}} P_\rho \otimes Q_\beta - Q_\beta\otimes \sigma^\rho_{\alpha\dot{\alpha}} P_\rho \right) \cr
&\cr
& - & \Psi^{\rho\alpha}\left(P_\rho \otimes \sigma^\lambda_{\alpha\dot{\alpha}} P_\lambda  -  \sigma^\lambda_{\alpha\dot{\alpha}} P_\lambda \otimes P_\rho \right) .
\label{cp3}
\eea
These comultiplication rules may again be used to verify the unmodified (anti-)commutation relations of the super-Poincar\'e algebra. Very recently, these rules were derived also from quantum group theoretic arguments \cite{Kobayashi:2004ep}. The comultiplication rules for the superconformal generators can also be found in a similar way. Especially, in the C-deformed superspace, we may use the differential operator representation in (\ref{deformscon}) to find the following rules for $\hat D, \hat A, \hat K_{\alpha\dot\alpha}, \hat S_\alpha$ and $\hat{\bar S}_{\dot\alpha}$ also:
\bea
\label{cp4}
\Delta(\hat D) & = &\hat D \otimes {\bf{1}} + {\bf{1}} \otimes \hat D
- \frac{i}{2} C^{\alpha\beta}\hat Q_\alpha\otimes \hat Q_\beta ,\cr
&\cr
\Delta(\hat A)& = &\hat A \otimes {\bf{1}} + {\bf{1}} \otimes \hat A - \frac{1}{2} C^{\alpha\beta}\hat Q_\alpha\otimes \hat Q_\beta ,\cr
&\cr
\Delta(\hat{{K}}_{\alpha\dot{\alpha}}) & = & \hat{{K}}_{\alpha\dot{\alpha}}\otimes{\bf{1}} + {\bf{1}} \otimes  \hat{{K}}_{\alpha\dot{\alpha}} + \epsilon_{\alpha\beta}C^{\beta\gamma}\left( \hat{\bar{S}}_{\dot{\alpha}}\otimes Q_\gamma + Q_\gamma \otimes  \hat{\bar{S}}_{\dot{\alpha}} \right) ,
\cr
&\cr
\Delta(\hat{S}_{\alpha}) & = & \hat{S}_{\alpha}\otimes{\bf{1}} + {\bf{1}} \otimes \hat{S}_{\alpha} \cr
&\cr
& + & i C^{\beta\gamma}\left( 2  \hat M_{\alpha\beta} +  \epsilon_{\alpha\beta}\hat D - 3 i \epsilon_{\alpha\beta}\hat A \right) \otimes Q_\gamma \cr
&\cr
& - & i C^{\beta\gamma} Q_\gamma \otimes \left(2  \hat M_{\alpha\beta} +  \epsilon_{\alpha\beta}\hat D - 3 i \epsilon_{\alpha\beta}\hat A  \right) \cr
&\cr
& + & i C^{~\gamma}_\alpha C^{\beta\delta} \left(\hat Q_\beta \otimes \hat  Q_\gamma \hat Q_\delta + \hat Q_\gamma \hat Q_\delta \otimes \hat Q_\beta\right) ,\cr
&\cr
\Delta(\hat{\bar{S}}_{\dot{\alpha}}) & = & \hat{\bar{S}}_{\dot{\alpha}}\otimes{\bf{1}} + {\bf{1}} \otimes \hat{\bar{S}}_{\dot{\alpha}} .
\eea
Up to our knowledge, these comultiplication rules have not been given before.

\section{Field transformations and symmetries}

The advantage of working with deformed generators is that they act as manifest symmetries of the equations of the motion in the non-(anti)commutative space. In this section we shall consider the transformation properties of the fields under conformal and super-Poincar\'e symmetries and the invariance of equations of motion as regards the deformed symmetries for some simple cases.  To discuss the transformation properties of the fields under conformal symmetry, we will here consider the generalization of the the ``minimal'' space-time representation (\ref{deformconrep}) in such a way that the differential operators $\hat\delta_d,\hat\delta_\omega $ and $\hat\delta_c$ may be given by the forms:
\bea
\label{genrep}
\hat \delta_d  &=&  -\hat x^\rho\hat\p_\rho - d,~~~~~~
\hat\delta_\omega  = -\hat x^\lambda \omega^{~\rho}_\lambda \hat\p_\rho + \frac{i}{2}\theta^{\mu\nu}\omega^{~\rho}_\mu\hat\p_\nu\hat\p_\rho + \frac{i}{2} \omega^{\mu\nu}\Sigma_{\mu\nu} \cr
&\cr
\hat\delta_c  &=& 2c.\hat{x}\hat x^\rho \hat\p_\rho - \hat x^2 c^\lambda\hat\p_\lambda - ic_\lambda\theta^{\lambda\rho}\hat x^\sigma\hat\p_\sigma\hat\p_\rho + ic^\lambda\theta^{\rho\sigma}\hat x^\rho\hat\p_\sigma\hat\p_\lambda\cr
&\cr
&& + \frac{1}{4} c_\lambda\theta^{\alpha\rho}\theta^{~\sigma}_\alpha \p_\rho\hat\p_\sigma\hat\p_\lambda + 2 d c.\hat x - i d c_\lambda\theta^{\lambda\rho}\hat\p_\rho - 2 i \hat x^\nu c^\lambda\Sigma_{\lambda\nu}\cr
&\cr
&& - c^\lambda\theta^{\nu\sigma}\Sigma_{\lambda\nu}\hat\p_\sigma.
\eea
These expressions differ from the minimal space-time representations by the presence of the spin matrix $\Sigma_{\mu\nu}$ (satisfying the Lorentz algebra by themselves) and the constant parameter $d$ which will assume the role of the scale dimension of the field in consideration. The differential operators in this generalized form also satisfy the commutation relation of the conformal algebra (\ref{boscon}), as one can easily verify. Then, the (deformed) conformal symmetry may be realized by taking the following transformations of the fields $\hat\psi(\hat x)$ (which can be Lorentz tensors/spinors with an appropriate choice of the spin matrix $\Sigma_{\mu\nu}$) : 
\be
\label{fieldtrans}
{\bs{\hat\delta}}^T \hat\psi = - a^\mu \hat\p_\mu \hat\psi,~~~{\bs{\hat\delta}}^L \hat\psi = - \hat\delta_\omega \hat\psi,~~~{\bs{\hat\delta}}^D \hat\psi = - \hat\delta_d \hat\psi,~~~{\bs{\hat\delta}}^C \hat\psi = - \hat\delta_c \hat\psi
\ee
The bold-faced ${\bs{\hat\delta}}$'s, acting on the fields $\hat\psi(\hat x)$ (with tensor/spinor indices suppressed), provide a 'field representation' of the conformal algebra in the non-commutative space and thus $({\bs{\hat\delta}}^T,{\bs{\hat\delta}}^L, {\bs{\hat\delta}}^D,{\bs{\hat\delta}}^C)$ satisfy the same commutation relations of the conformal algebra as the space-time representations  $(\hat\delta_a\equiv a^\mu\hat\p_\mu,\hat\delta_\omega, \hat\delta_d,\hat\delta_c)$ do.

If the field $\hat\psi(\hat x)$ is taken to be an elementary scalar field $\hat\phi(\hat x)$ in 4-dimensional space-time, one may choose $\Sigma_{\mu\nu} = 0$ and $d=1$ in (\ref{fieldtrans}). Then, for the related derivatives fields $\hat\p_\mu\hat\phi(\hat x)$ and $\hat\p^2 \hat \phi (\hat x)$, one may use (\ref{fieldtrans}) to derive the appropriate transformation properties. For the field $\hat V_\mu \equiv \hat\p_\mu\hat\phi(\hat x)$, one finds that their Poincar\'e and scale transformations are consistent with our field transformation laws in (\ref{fieldtrans}) only if we now take the Lorentz-vector spin matrix $(\Sigma^{(V)}_{\mu\nu})^{~\lambda}_{\rho} = i (\eta_{\mu\rho}\eta^{~\lambda}_\nu - \eta_{\nu\rho}\eta^{~\lambda}_\mu )$ and the scale dimension $d = 2$ for $\hat V_\mu (\hat x)$: explicitly using the notation like $\delta^{(0)}_d,\delta^{(0)}_\omega, $ etc. to indicate the operators obtained with $\Sigma_{\mu\nu} = 0$ and $ d = 1$ in (\ref{genrep}),
\bea
{\bs{\hat\delta}}^T (\hat V_\rho) &\equiv& \hat\p_\rho ({\bs{\hat\delta}}^T \hat\phi) = - a^\mu\hat\p_\mu(\hat\p_\rho\hat\phi) =  - a^\mu\hat\p_\mu\hat V_\rho,\cr
&\cr
{\bs{\hat\delta}}^L (\hat V_\rho) &\equiv& \hat\p_\rho ({\bs{\hat\delta}}^L \hat\phi) = - \hat\delta^{(0)}_\omega (\hat\p_\rho\hat\phi) - \left[\hat\p_\rho, \hat\delta^{(0)}_\omega\right]\hat\phi\cr
&\cr
&=& -\hat\delta^{(0)}_\omega(\hat\p_\rho\hat\phi) + \omega^{~\lambda}_\rho (\hat\p_\lambda\hat\phi)~\equiv - \hat\delta^{(V)}_\omega (\hat V_\rho),\cr
&\cr
{\bs{\hat\delta}}^D(\hat V_\rho) &\equiv& \hat\p_\rho ({\bs{\hat\delta}}^D \hat\phi) = - \hat\delta^{(0)}_d (\hat\p_\rho\hat\phi) - \left[\hat\p_\rho, \hat\delta^{(0)}_d\right]\hat\phi\cr
&\cr
&=& -\left(-\hat x^\nu\hat\p_\nu - 2 \right)\hat V_\rho.
\eea
But, under special conformal transformations, $\hat V_\mu (\hat x)$ can not be described by our formulas in (\ref{fieldtrans}). In contrast, for the field $\hat\chi(\hat x) \equiv \hat\p^2\hat\phi(\hat x)$, one finds after some straightforward algebra that its full conformal transformation behaviors, including those  under special conformal transformations, are described by our formulas (\ref{fieldtrans}) (appropriate to a {\it{dimension-3}} Lorentz scalar field $\hat\chi(\hat x)$).

We now turn to the action of these symmetry generators on the local product of the fields. This can be done by making use of the comultiplication structure derived in (\ref{cp0})-(\ref{coproduct2}). Here it should be noted that, even if more general differential operators in (\ref{genrep}) are allowed to act on product of two functions $\hat f\hat g$ (with different sets of $\Sigma_{\mu\nu}$ and $d$ chosen for $\hat f$ and $\hat g$), the comultiplication rules as given by the forms in (\ref{cp0})-(\ref{coproduct2}) continue to satisfy the commutation relation of the conformal algebra. Based on this bialgebra structure, we may now take the conformal transformations for the composite fields $\hat\psi_1\otimes\hat\psi_2$ to be given by
\be
\label{comptrans}
{\bs{\hat\delta}}^X \hat\psi_1\otimes\hat\psi_2 = - \Delta(\hat \delta_x)\hat\psi_1\otimes\hat\psi_2,
\ee
where $\Delta(\hat \delta_x) $ on the right denotes the differential operator acting on  $\hat\psi_1\otimes\hat\psi_2$ according to the rules (\ref{cp0})-(\ref{coproduct2}). Here the superscript $X$ (subscript $x$) can refer to any of the full conformal transformations. 

Using (\ref{comptrans}), it is not difficult to derive the conformal transformation properties of the composite fields like $\hat\Psi_1(\hat x) \equiv \hat\phi^2(\hat x)$ and $\hat\Psi_2(\hat x) \equiv \hat\phi^3(\hat x)$, where $\hat\phi(\hat x)$ is a elementary, dimension-1, scalar field with its transformation properties specified as above. The results are simply
\bea
\hat {\bs{\delta}}^X \hat\Psi_1 &=& -\hat \delta_x \hat\Psi_1 ~({\rm with} ~ \Sigma_{\mu\nu} = 0 ~{\rm and}~ d = 2 ~for~ \hat\Psi_1)\cr
&\cr
\hat {\bs{\delta}}^X \hat\Psi_2 &=& -\hat \delta_x \hat\Psi_2 ~({\rm with} ~ \Sigma_{\mu\nu} = 0 ~{\rm and}~ d = 3 ~for~ \hat\Psi_2)
\eea
i.e., despite the twisted comultiplication, their conformal transformations are exactly of the same form as those of an elementary field. Use of the higher derivative representation (\ref{genrep}) for $\hat\delta_x$ is responsible for this. Furthermore, as the fields $\hat\p^2 \hat\phi (\hat x)$ and $\hat\phi^3(\hat x)$ have identical conformal transformation properties (in 4-dimensional space-time), we can conclude that the scalar field equation of the form
\be
\hat\p^2 \hat \phi + g \hat\phi^3 = 0.
\ee
is invariant under full (deformed) conformal transformations. This is the generalization of the well known fact for a massless $\phi^4$-theory in ordinary space-time.

Let us now consider the transformation properties of the superfields in the non-(anti)commutative superspace. For simplicity, only the super-Poincar\'e transformations on scalar superfields will be considered here. Though the supercharges $\hat{\bar Q}_{\dot\alpha}$ are deformed in the non-(anti)commutative superspace, the covariant derivatives  remain unchanged
\bea
\hat D_\alpha    =  i \hat{\p}_{\alpha} - 2 \hat{\bar\theta}^{\dot\alpha} \hat\p_{\alpha \dot{\alpha}}, ~~~~~~~~
\hat{\bar{D}}_{\dot{\alpha}}  =  - i\hat{\bar\p}_{\dot\alpha}~.
\eea
One can define the chiral superfield $\hat \Phi (\hat y,\hat \theta)$ in the usual manner, i.e., by demanding that $\hat{\bar D}_{\dot\alpha}\hat\Phi (y,\theta) = 0 $. For component expansion in $\hat\theta$, we can write it as
\bea
\hat\Phi (\hat y,\hat \theta) & =& \hat a(\hat y) + \sqrt{2}\hat\theta\hat\psi(\hat y) + \hat \theta^2 \hat f(\hat y).
\eea
The anti-chiral superfield $\hat\Phi (\hat{\bar y},\hat{\bar \theta})$ can be expanded analogously by rewriting first, $\hat{\bar y}$ in terms of $\hat y$, etc.

To write the supersymmetry transformation of the chiral superfields, we follow the approach similar to the bosonic symmetries and write the field transformations as
\bea
{\bs{\hat\delta}}^\epsilon \hat \Phi &=& - \hat \delta_\epsilon  \hat \Phi ~= -i \epsilon^\alpha(i\hat\p_\alpha)\hat\Phi,\cr
&\cr
{\bs{\hat\delta}}^{\bar \epsilon} \hat\Phi &=& - \hat\delta_{\bar\epsilon} \hat \Phi ~= - i \bar\epsilon^{\dot\alpha}(- i \hat{\bar{\p}}_{\dot{\alpha}} -2 \hat\theta^\alpha \hat\p_{\alpha \dot{\alpha}} + C^{\alpha\beta}\hat\p_\beta \hat\p_{\alpha\dot{\alpha}} - \Psi^{\rho\alpha}\hat\p_\rho \hat\p_{\alpha\dot{\alpha}})\hat\Phi.
\eea
Then for the covariant derivatives of the chiral superfield, one finds
\bea
{\bs{\hat\delta}}^\epsilon (\hat D_\alpha \hat \Phi) &\equiv& \hat D_\alpha ({\bs{\hat\delta}}^\epsilon\hat\Phi) ~= - \hat\delta_\epsilon(\hat D_\alpha \hat \Phi) -\left[\hat D_\alpha, \hat\delta_\epsilon\right] ~=  - \hat\delta_\epsilon(\hat D_\alpha \hat \Phi), \cr
&\cr
{\bs{\hat\delta}}^{\bar \epsilon} (\hat D_\alpha \hat\Phi) &\equiv& \hat D_\alpha ({\bs{\hat\delta}}^{\bar \epsilon} \hat\Phi) ~= - \hat\delta_{\bar\epsilon}(\hat D_\alpha \hat \Phi) -\left[\hat D_\alpha, \hat\delta_{\bar\epsilon}\right] ~= - \hat\delta_{\bar\epsilon}(\hat D_\alpha \hat \Phi).
\eea
The transformations rules for the anti-chiral superfield and the corresponding covariant derivatives can be written in a similar fashion. Using the above relations and acting with $\hat D_\alpha$ ( $\hat {\bar D}_\alpha$) once more, it is an easy exercise to show that $\hat D^2 \hat \Phi$ ($ {\hat {\bar D}}^2 \hat {\bar\Phi} $) transform as a chiral (anti-chiral) superfield under the deformed supersymmetry.

For the local product of two (or more) chiral superfields, the transformations rules can be written using the comultiplication rules (\ref{comsusy}). Using the bialgebra structure of the comultiplication rules (\ref{comsusy}), the transformation of the product $\hat\Phi_1 \otimes\hat\Phi_2$ can be taken by the formulas 
\be
{\bs{\hat\delta}}^\epsilon \hat\Phi_1 \otimes\hat\Phi_2 = - \Delta(\hat \delta_\epsilon)\hat\Phi_1 \otimes\hat\Phi_2,~~~
{\bs{\hat\delta}}^{\bar \epsilon}\hat\Phi_1\otimes\hat\Phi_2 = - \Delta(\hat {\delta}_{\bar\epsilon})\hat\Phi_1\otimes\hat\Phi_2.
\ee
This has the simple consequence that the transformation law for the product of two chiral superfields, $\hat\Psi \equiv \hat \Phi^2$ behaves again as that of a chiral superfield, i.e.,
\be
{\bs{\hat\delta}}^\epsilon \hat\Psi = -\hat\delta_\epsilon\hat\Psi, ~~~{\bs{\hat\delta}}^{\bar \epsilon}\hat\Psi = - \hat{\delta}_{\bar\epsilon}\hat\Psi.
\ee
Note that the higher-derivative nature of $\hat\delta_{\bar\epsilon}$ and the related modified comultiplication rule $\Delta(\hat{\bar Q}_{\dot\alpha})$ in (\ref{cp3}) are crucial to ensure this behavior. As the supersymmetry transformations of  $ {\hat {\bar D}}^2 \hat {\bar\Phi} $ and $\hat\Phi^2$ are exactly the same as that of a chiral superfield, we can immediately conclude for instance  that equation of motion for the Wess-Zumino model 
\be
\hat{\bar D}^2\hat{\bar\Phi} - m\hat\Phi - g \hat\Phi^2 = 0
\ee
is manifestly invariant under deformed supersymmetry. The above discussion can be generalized to other superfields (e.g. vector superfields) and one can write the corresponding operator equations in the non-(anti)commutative superspace which are invariant under deformed supersymmetry.

\section{Concluding remarks}

In this work we have considered the differential calculus in the deformed superspace, generalizing the corresponding approaches \cite{Wess:2003da, Dimitrijevic:2003wv} in the deformed Minkowski space. Based on this, we have found the differential operator representations in the deformed space for the conformal, super-Poincar\'e and superconformal generators. They contain higher-derivative terms, but satisfy the same (anti-)commutation relations of the algebras as in the undeformed case. The higher derivative terms are responsible for the modifications in the comultiplication structures and we have exhibited them explicitly. We then considered the transformation of the fields under deformed symmetry and exhibited that they can provide a natural realization of the field theoretic symmetry in the deformed spaces. Though the symmetry can not be given a meaning in the sense of the Noether's theorem because of the ``twisted'' comultiplication structure used to derive the transformation laws for the composite fields, it serves a useful tool to write the invariant equations of motion in the deformed spaces. We hope that the differential calculus developed in this paper and deformed symmetry consideration can be useful for studying various aspects of field theories defined on non-(anti)commutative superspace. It is desirable to study for instance the consequences of these symmetries on soliton solutions and on issues related to quantum loop corrections such as the renormalization and anomaly structures of field theories defined in the noncommutative spaces. One may also study the supergravity theories in the deformed superspace using this formalism. Some of these issues are under consideration. 

\vskip .2in

{\bf Acknowledgments:} The work of CL and SS was supported by Korea Science Foundation ABRL program (R14-2003-012-01001-0). One of the authors (RB) would like to thank the Centre for Theoretical Physics, Seoul National University, where this work was initiated, for providing support during his stay.



\begin{thebibliography}{99}

\bibitem{Snyder}
H. S. Snyder, Phys. rev. 71 (1947) 38; Phys. Rev. 72 (1947) 68.

\bibitem{Dunne:1989hv}
  G.~V.~Dunne, R.~Jackiw and C.~A.~Trugenberger,
  Phys.\ Rev.\ D {\bf 41} (1990) 661.

\bibitem{Seiberg:1999vs}
  N.~Seiberg and E.~Witten,
  JHEP\ {\bf 9909} (1999) 032
  [arXiv:hep-th/9908142].


\bibitem{Douglas:2001ba}
  M.~R.~Douglas and N.~A.~Nekrasov,
  Rev.\ Mod.\ Phys.\  {\bf 73} (2001) 977
  [arXiv:hep-th/0106048];
  E.~Langmann, R.~J.~Szabo and K.~Zarembo,
  JHEP {\bf 0401} (2004) 017
  [arXiv:hep-th/0308043].

\bibitem{Moyal}
J. E.Moyal, Proc. Cambridge Phil. Soc. 45 (1949) 99; H. J. Groenewold, Physica 12 (1946) 405; H.Weyl, Z. Phys. 46 (1927) 1.

\bibitem{Chaichian:2004za}
  M.~Chaichian, P.~P.~Kulish, K.~Nishijima and A.~Tureanu,
  Phys.\ Lett.\ B {\bf 604} (2004) 98
  [arXiv:hep-th/0408069];
  M.~Chaichian, P.~Presnajder and A.~Tureanu,
  Phys.\ Rev.\ Lett.\  {\bf 94} (2005) 151602
  [arXiv:hep-th/0409096].

\bibitem{Wess:2003da}
  J.~Wess,
  arXiv:hep-th/0408080;
  M.~Dimitrijevic and J.~Wess,
  arXiv:hep-th/0411224.

\bibitem{Dimitrijevic:2003wv}
  M.~Dimitrijevic, L.~Jonke, L.~Moller, E.~Tsouchnika, J.~Wess and M.~Wohlgenannt,
  Eur.\ Phys.\ J.\ C {\bf 31} (2003) 129
  [arXiv:hep-th/0307149];
  M.~Dimitrijevic, L.~Moller and E.~Tsouchnika,
  J.\ Phys.\ A {\bf 37} (2004) 9749
  [arXiv:hep-th/0404224];
  M.~Dimitrijevic, L.~Jonke, L.~Moller, E.~Tsouchnika, J.~Wess and M.~Wohlgenannt,
  Czech.\ J.\ Phys.\  {\bf 54} (2004) 1243
  [arXiv:hep-th/0407187].

\bibitem{Rabin}
  R.~Banerjee,
  arXiv:hep-th/0508224.

\bibitem{Ferrara:2000mm}
  S.~Ferrara and M.~A.~Lledo,
  JHEP {\bf 0005} (2000) 008
  [arXiv:hep-th/0002084];
  D.~Klemm, S.~Penati and L.~Tamassia,
  Class.\ Quant.\ Grav.\  {\bf 20} (2003) 2905
  [arXiv:hep-th/0104190];
  J.~de Boer, P.~A.~Grassi and P.~van Nieuwenhuizen,
  Phys.\ Lett.\ B {\bf 574} (2003) 98
  [arXiv:hep-th/0302078].

\bibitem{Seiberg:2003yz}
  N.~Seiberg,
  JHEP {\bf 0306} (2003) 010
  [arXiv:hep-th/0305248].




\bibitem{Koch:2004ud}
  F.~Koch and E.~Tsouchnika,
  Nucl.\ Phys.\ B {\bf 717} (2005) 387
  [arXiv:hep-th/0409012].

\bibitem{Matlock:2005zn}
  P.~Matlock,
  Phys.\ Rev.\ D {\bf 71} (2005) 126007
  [arXiv:hep-th/0504084].

\bibitem{wess}
J. Wess and J. Bagger, {\it Supersymmetry and Supergravity}, (Princeton University Press, Princeton, 1992).


\bibitem{buchbinder}
I. L. Buchbinder, S. M. Kuzenko, {\it Ideas and Methods of Supersymmetry and Supergravity or a Walk Through Superspace}, (Institute of Physics Publishing, Bristol and Philadelphia, 1995).

\bibitem{Kobayashi:2004ep}
  Y.~Kobayashi and S.~Sasaki,
  Int.\ J.\ Mod.\ Phys.\ A {\bf 20} (2005) 7175
  [arXiv:hep-th/0410164];
  B.~M.~Zupnik,
  Phys.\ Lett.\ B {\bf 627} (2005) 208
  [arXiv:hep-th/0506043];
  M.~Ihl and C.~Saemann,
  JHEP {\bf 0601} (2006) 065
  [arXiv:hep-th/0506057].

\end{thebibliography}
\end{document}